\documentclass{iopart}

\usepackage{times}
\usepackage{graphicx}
\usepackage{iopams}

\renewcommand{\vec}[1]{\boldsymbol{#1}}

\begin{document}

%%% TITLE %%%%%%%%%%%%%%%%%%%%%%%%%%%%%%%%%%%%%%%%%%%%%%%%%%%%%%%%%%%%%%%%%%%%%

\title{Rotational Diffusion in a Chain of Particles}
\author{Holger Stark\dag, Michael Reichert\dag, and J\'erome Bibette\ddag}
\address{\dag\ Universit\"at Konstanz, Fachbereich Physik,
               D-78457 Konstanz, Germany}
\address{\ddag\ Laboratoire Colloides et Mat\'eriaux Divis\'es, UMR 7612,
                ESPCI, 10 rue Vauquelin, 75005 Paris, France}

\ead{Holger.Stark@uni-konstanz.de}

\begin{abstract}
We study the coupled rotational diffusion in a two-particle chain
on the basis of a Smoluchowski equation and calculate time-correlation
functions that are measurable in an experiment. This might be used
to explore hydrodynamic interactions in the limit where lubrication
theory is valid.
\end{abstract}


%%% INTRODUCTION %%%%%%%%%%%%%%%%%%%%%%%%%%%%%%%%%%%%%%%%%%%%%%%%%%%%%%%%%%%%%%

\section{Introduction} \label{sec:introduction}

Colloids are widely considered as models for atomic systems\ 
\cite{Pusey1991,Poon1996}. 
%However, their hydrodynamic interactions 
%are a specific signature of colloidal suspensions
However, a specific signature of colloidal suspensions are hydrodynamic 
interactions\ \cite{Happel73,Dhont96}.
Moving colloids interact with each other through
the flow fields they create. This
%The interaction of colloids through the flow fields which they create
is a true multi-body problem which can only be handled by approximate
methods such as multipole expansions for large particle distances 
(see, e.g., Refs.\ \cite{Dhont96,Cichocki94})
and lubrication theory when they come close to each other\ \cite{Kim91}.
Conventionally, hydrodynamic interactions are monitored through their
effect on self and collective diffusion in colloidal suspensions\ 
\cite{Pusey1991,Dhont96,Naegele96} but recent experiments with optical 
tweezers on a pair of particles\ \cite{fewparticles}
also allow a controlled exploration of hydrodynamic interactions as 
a function of particle separation confirming standard approaches 
due to Oseen and Rotne-Prager\ \cite{Happel73,Dhont96}.

Recent work also studied the rotational diffusion of tracer particles\ 
\cite{Montgomery77,tracer,Koenderink01}
or colloids trapped in optical tweezers\ \cite{Reichert04}. An experimental
system introduced by Bibette {\em et al.} \cite{Bibette} suggests a 
possibility to directly measure the effect of hydrodynamic interactions 
on the rotational diffusion in the limit where lubrication theory is valid.
Charged superparamagnetic particles under the influence of a magnetic
field form chains where the particle separation can precisely be tuned by the
magnetic field strength. By labeling the particles with a phosphorescent
dye\ \cite{Koenderink01} or by using birefringent colloids\ \cite{Mertelj02}, 
rotational diffusion can then be monitored. 

The present article investigates 
rotational diffusion in a two-particle chain theoretically. It first 
introduces the Smoluchowski equation to treat rotational diffusion
and then discusses observabels to be measured in an experiment.

\section{Rotational Diffusion and the Smoluchowski Equation} \label{sec:smolu}

Let us first shortly review the rotational diffusion of one particle
that can also be considered as a random walk on the unit sphere.
The probability density $P(\hat{\vec{\nu}},t)$ of finding the particle
with an orientation given by the unit vector $\hat{\vec{\nu}}$ at time
$t$ satisfies the Smoluchowski equation\ \cite{Montgomery77}
\begin{equation}
\frac{\partial P(\hat{\vec{\nu}},t)}{\partial t} = D_{0} \vec{\nabla}_{r}^{2}
P(\hat{\vec{\nu}},t) \enspace.
\label{01}
\end{equation}
This is in complete analogy with translational diffusion. However,
instead of the Laplace operator, the square of the nabla operator in 
angular space $\vec{\nabla}_{r}$ has to be used, where the index $r$
means rotation. Since 
$\vec{\nabla}_{r} = \frac{2\pi i}{h} \vec{L}$, where $\vec{L}$ is the 
angular momentum operator known from quantum mechanics, all the
algebra developed for $\vec{L}$ \cite{Messiah61}
is also valid for $\vec{\nabla}_{r}$.
To be concrete, we note that
\begin{equation}
\vec{\nabla}_{r} = \hat{\vec{\nu}} \times 
\frac{\partial}{\partial \hat{\vec{\nu}}} =
\left( \begin{array}{c}
         -\sin \varphi \frac{\partial}{\partial \vartheta} - 
          \frac{\cos \varphi}{\tan \vartheta} 
          \frac{\partial}{\partial \varphi} \\
         -\cos \varphi \frac{\partial}{\partial \vartheta} - 
          \frac{\sin \varphi}{\tan \vartheta} 
          \frac{\partial}{\partial \varphi} \\
          \frac{\partial}{\partial \varphi}
       \end{array}
\right) \enspace,
\label{1}
\end{equation}
where $\phi,\vartheta$ are the spherical coordinates to represent 
$\hat{\vec{\nu}}$. The rotational diffusion constant $D_{0}$ in 
Eq.\ (\ref{01}) is related via an Einstein relation to the mobility 
$\mu_{0} = 1/(8\pi \eta a^3)$ for rotational motion, i.e.,
$D_{0} = k_B T \mu_{0}$. Now, since the operator $\vec{\nabla}_{r}^{2}$
possesses the spherical harmonics $Y^{l}_{m}(\hat{\vec{\nu}})$ as its
eigenfunctions, $\vec{\nabla}_{r}^{2}Y^{l}_{m}(\hat{\vec{\nu}}) =
-l(l+1) Y^{l}_{m}(\hat{\vec{\nu}})$, one calculates, with the same method
as presented below, the time-correlation functions
\begin{equation}
\langle \, Y^{l*}_{m}[\hat{\vec{\nu}}(t)] \, 
Y^{l^{\prime}}_{m^{\prime}}[\hat{\vec{\nu}}(0) ] \, \rangle = \frac{1}{4\pi}
\delta_{ll^{\prime}} \delta_{mm^{\prime}} e^{-l(l+1)D_0 t} \enspace.
\label{1a}
\end{equation}
Here $\hat{\vec{\nu}}(t)$ means orientation of the particle at
time $t$ and the symbol $*$ means complex conjugate. For small times,
one shows with the help of Eq. (\ref{1a}) (for details, see below) 
that the square
of the angular displacement of $\hat{\vec{\nu}}$ exhibits the typical
diffusive behavior:
\begin{equation}
\langle \, |\hat{\vec{\nu}}(t) - \hat{\vec{\nu}}(0)|^{2} \, \rangle
\approx 4 D_0 t \enspace.
\label{1b}
\end{equation}

\begin{figure}
%\begin{indented}
%\item
\includegraphics[angle=-90,width=0.95\textwidth]{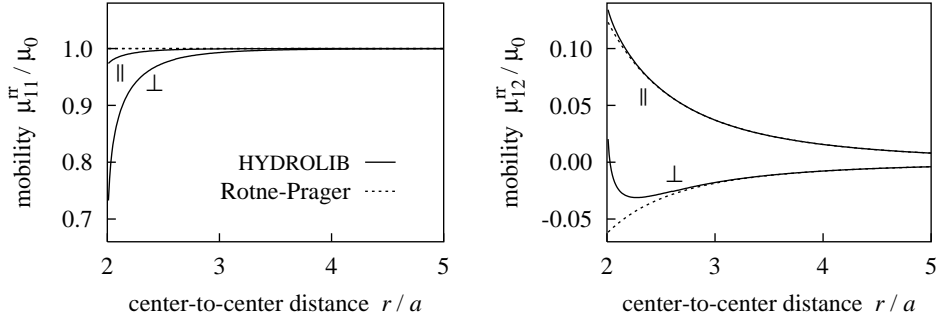}
%\end{indented}
\caption{The rotational mobilities of the two-particle chain in units
  of $\mu_0=1/(8 \pi \eta a^3)$ as a function of reduced 
  center-to-center distance $r/a$. Full lines: values calculated with
  numerical library HydroLib; dotted lines: the Rotne-Prager approximation.}
\label{fig:mobi}
\end{figure}

We now concentrate on a chain of two particles of radius $a$ whose
centers are connected by the vector $\vec{r}$. For simplicity, we only
consider their rotational diffusion, i.e., we disregard any coupling
to translational motion. We especially take the particles' separation
$r$ as fixed, e.g., by assuming that fluctuations around the
equilibrium separation, governed by the two-particle potential, are 
negligibly small. 
The coupled rotational diffusion is described by self-diffusion tensors
$\vec{D}_{11} = \vec{D}_{22}$ for particle 1 and 2, respectively, and 
the tensors $\vec{D}_{12} = \vec{D}_{21}$ encoding
the hydrodynamic interactions between particle 1 and 2.
As in the one-particle case, these quantities are related to
mobilities by an Einstein relation, $\vec{D}_{ij} = k_{B}T \vec{\mu}_{ij}$.
The mobilites $\vec{\mu}_{ij}$ connect the torque $\vec{T}_{j}$ on particle $j$
with the angular velocity $\vec{\omega}_{i}$ of particle $i$:
$\vec{\omega}_{i} = \vec{\mu}_{ij} \vec{T}_{j}$. Due to Lorentz's
reciprocal theorem\ \cite{Happel73}, they fulfill 
$\vec{\mu}_{ij} = \vec{\mu}_{ji}^{t}$, where $t$ means transposed matrix.
If, in addition, the particles are identical, 
$\vec{\mu}_{ij} = \vec{\mu}_{ji}$. The uniaxial
symmetry of the two-particle chain determines the form of $\vec{D}_{ij}$:
\begin{equation}
\vec{D}_{ij} = D_{ij}^{\perp} \vec{1} + \Delta D_{ij} 
\hat{\vec{r}} \otimes \hat{\vec{r}} \enspace \mathrm{with} \enspace 
\Delta D_{ij} = D_{ij}^{\|} - D_{ij}^{\perp} \enspace,
\label{2}
\end{equation}
where $\vec{1}$ means unit tensor, $\hat{\vec{r}} = \vec{r}/r$, and
$\otimes$ means tensor product. The constants $D_{ij}^{\|}$ and 
$D_{ij}^{\perp}$ refer, respectively, to rotational diffusion about 
the two-particle axis $\hat{\vec{r}}$ or a direction perpendicular to it.
The related mobilities as a function of reduced particle distance $r/a$
are plotted in Fig.\ \ref{fig:mobi}. The library HydroLib\ \cite{Hinsen95} 
allows to 
calculate their values ranging from small particle distances where
lubrication theory has to be applied to large distances where expansions
into $a/r$ are applicable. For comparison, the Rotne-Prager approximation
is also shown. Both graphs of Fig.\ \ref{fig:mobi} illustrate that the 
Rotne-Prager approximation works well for center-to-center distances
larger than $3a$. To see noticable deviations of the mobilities
$\mu_{11}^{rr\|}$ and $\mu_{11}^{rr\perp}$ from the single-particle
value $\mu_0=1/(8 \pi \eta a^3)$ (the index $rr$ refers directly to
the rotational degree of freedom), the particles have to be close.
Note that the mobilities in both graphs stay finite when the particles
approach contact at $r=2a$; the derivatives of the perpendicular
coefficients, however, are singular.

The Smoluchowski equation determines the temporal evolution of the
probability density $P(\hat{\vec{\nu}}_{1},\hat{\vec{\nu}}_{2},t)$
to find particles 1 and 2 with respective orientations given by
$\hat{\vec{\nu}}_{1}$ and $\hat{\vec{\nu}}_{2}$ at time $t$\ 
\cite{Dhont96,Montgomery77}:
\begin{equation}
\frac{\partial P(\hat{\vec{\nu}}_{1},\hat{\vec{\nu}}_{2},t)}{\partial t} 
= \hat{\vec{L}}_{S} P(\hat{\vec{\nu}}_{1},\hat{\vec{\nu}}_{2},t),
\label{3}
\end{equation}
where 
\begin{equation}
\hat{\vec{L}}_{S} = \vec{\nabla}_{r1} \cdot \vec{D}_{11} \vec{\nabla}_{r1} 
+ \vec{\nabla}_{r2} \cdot \vec{D}_{22} \vec{\nabla}_{r2}
+ 2 \vec{\nabla}_{r1} \cdot \vec{D}_{12} \vec{\nabla}_{r2}
\label{4}
\end{equation}
denotes the Smoluchowski operator that is a generalization of 
$D_{0} \vec{\nabla}_{r}^{2}$ in the single-particle equation (\ref{01}).
Since both particles are identical and with the help of Eq. (\ref{2}), it is
rewritten as
\begin{eqnarray}
\hat{\vec{L}}_{S} & = & D_{11}^{\perp} (\vec{\nabla}_{r1}^{2} + 
\vec{\nabla}_{r2}^{2}) + \Delta D_{11} 
\Big(\frac{\partial^{2}}{\partial \varphi_{1}^{2}} + 
     \frac{\partial^{2}}{\partial \varphi_{2}^{2}}\Big) \nonumber\\
& & + 2 D_{12}^{\perp} \vec{\nabla}_{r1} \cdot \vec{\nabla}_{r2}
+ 2 \Delta D_{12} \frac{\partial}{\partial \varphi_{1}}
    \frac{\partial}{\partial \varphi_{2}}
\label{5}
\end{eqnarray}
where $\partial/\partial \varphi_{i}$ is the $z$ component of 
$\vec{\nabla}_{ri}$ with $\hat{\vec{z}} \| \vec{r}$. 
In analogy to the wave function determined from 
Schr\"odinger's equation in quantum mechanics, the time evolution of the
probability density $P(\hat{\vec{\nu}}_{1},\hat{\vec{\nu}}_{2},t)$
is known in principle when eigenvalues and eigenfunctions of the 
Smoluchowski operator are known. Due to the analogy with the
angular-momentum algebra, the eigenvectors of the unperturbed problem,
i.e., particles not coupled by hydrodynamic interactions 
($D_{12}^{\perp}=\Delta D_{12}=0$), are just products of two spherical 
harmonics: 
$\Phi(\hat{\vec{\nu}}_{1},\hat{\vec{\nu}}_{2}) = 
Y^{l_{1}}_{m_{1}}(\hat{\vec{\nu}}_{1}) Y^{l_{1}}_{m_{2}}(\hat{\vec{\nu}}_{2})$.
They are even eigenfunctions of the last term in the second line of
Eq.\ (\ref{5}). The operator $\vec{\nabla}_{r1} \cdot \vec{\nabla}_{r2}$, 
however, mixes the
eigenfunctions of the unperturbed problem. How this is done, can be
calculated with the help of ``ladder operators'' introduced in full
analogy to the angular momentum algebra\ \cite{Messiah61}:
\begin{equation}
\nabla_{r}^{\pm} = \nabla_{rx} \pm i \nabla_{ry}
\end{equation}
Applied to spherical harmonics, they yield
\begin{equation}
\nabla_{r}^{\pm} Y^{l}_{m}(\hat{\vec{\nu}}) =
i \sqrt{l(l+1)-m(m \pm 1)} Y^{l}_{m \pm 1}(\hat{\vec{\nu}})
= i c^{\pm}_{lm} Y^{l}_{m \pm 1}(\hat{\vec{\nu}}) \enspace.
\label{6}
\end{equation}
We rewrite the crucial term $\vec{\nabla}_{r1} \cdot \vec{\nabla}_{r2}$
of $\hat{\vec{L}}_{S}$ with the help of
\begin{equation}
\nabla_{rx} = \frac{1}{2}(\nabla^{+}_{r} + \nabla^{-}_{r}) 
\enspace \mathrm{and} \enspace 
\nabla_{ry} = \frac{1}{2i}(\nabla^{+}_{r} - \nabla^{-}_{r})
\label{7}
\end{equation}
as
\begin{equation}
\vec{\nabla}_{r1} \cdot \vec{\nabla}_{r2} = \frac{1}{2}
(\nabla^{+}_{r1} \nabla^{-}_{r2} + \nabla^{-}_{r1} \nabla^{+}_{r2})
+ \frac{\partial }{\partial \varphi_{1}} \frac{\partial }{\partial \varphi_{2}}
%\enspace.
\label{8}
\end{equation}
and can now fully determine how the Smoluchowski operator acts on the
unperturbed eigenfunction:
\begin{eqnarray}
\fl
-\hat{\vec{L}}_{S} Y^{l_{1}}_{m_{1}} Y^{l_{2}}_{m_{2}} & = & 
\{ D^{\perp}_{11} [l_{1}(l_{1}+1)+l_{2}(l_{2}+1)]
+ \Delta D_{11} (m_{1}^{2}+m_{2}^{2}) + 2 D^{\|}_{12} m_{1}m_{2} \}
Y^{l_{1}}_{m_{1}} Y^{l_{2}}_{m_{2}}
\nonumber \\
\fl
& & + D^{\perp}_{12} (c^{+}_{l_{1}m_{1}} c^{-}_{l_{2}m_{2}} 
                      Y^{l_{1}}_{m_{1}+1} Y^{l_{2}}_{m_{2}-1}
    +                 c^{-}_{l_{1}m_{1}} c^{+}_{l_{2}m_{2}} 
                      Y^{l_{1}}_{m_{1}-1} Y^{l_{2}}_{m_{2}+1})
\label{9}
\end{eqnarray}
where we used $D^{\|}_{12} = D^{\perp}_{12} + \Delta D_{12}$ and the
coefficients $c^{\pm}_{lm}$ are defined in Eq.\ (\ref{6}).

\section{Time Correlation Functions and Observables} \label{sec:corr}

We are interested in quantities that can be measured in experiments.
We therefore define the most general time correlation function
\begin{equation}
\Gamma(^{l_{1}l_{2}m_{1}m_{2}}_{l_{1}^{{\prime}}l_{2}^{\prime}
            m_{1}^{\prime}m_{2}^{\prime}} | t ) =
\langle \, Y^{l_{1}*}_{m_{1}}[\hat{\vec{\nu}}_{1}(t)] \,
           Y^{l_{2}*}_{m_{2}}[\hat{\vec{\nu}}_{2}(t)] \,\, 
           Y^{l_{1}^{\prime}}_{m_{1}^{\prime}}[\hat{\vec{\nu}}_{1}(0)] \,  
           Y^{l_{2}^{\prime}}_{m_{2}^{\prime}}[\hat{\vec{\nu}}_{2}(0)]  
\,\rangle
\label{10}
\end{equation}
which can be related to useful observables. Formally, $\Gamma$ is calculated 
using the propagator 
$P(\hat{\vec{\nu}}_{1}, \hat{\vec{\nu}}_{2},t|
   \hat{\vec{\nu}}_{1}^{\prime}, \hat{\vec{\nu}}_{2}^{\prime},0)$ that
gives the probability of finding the two particles with orientations
$\hat{\vec{\nu}}_{1}$ and $\hat{\vec{\nu}}_{2}$ at time $t$ when they had with
certainty the orientations $\hat{\vec{\nu}}_{1}^{\prime}$ and 
$\hat{\vec{\nu}}_{2}^{\prime}$ at time $t=0$:
\begin{eqnarray}
\Gamma(^{l_{1}l_{2}m_{1}m_{2}}_{l_{1}^{{\prime}}l_{2}^{\prime}
            m_{1}^{\prime}m_{2}^{\prime}} | t ) & = &
\int\int Y^{l_{1}*}_{m_{1}}(\hat{\vec{\nu}}_{1})
         Y^{l_{2}*}_{m_{2}}(\hat{\vec{\nu}}_{2}) \,
         P(\hat{\vec{\nu}}_{1}, \hat{\vec{\nu}}_{2},t|
   \hat{\vec{\nu}}_{1}^{\prime}, \hat{\vec{\nu}}_{2}^{\prime},0) \, 
\label{11} \\
    & &  \times 
         W(\hat{\vec{\nu}}_{1}^{\prime}, \hat{\vec{\nu}}_{2}^{\prime},t=0)\,
         Y^{l_{1}^{\prime}}_{m_{1}^{\prime}}(\hat{\vec{\nu}}_{1}^{\prime}) \,  
           Y^{l_{2}^{\prime}}_{m_{2}^{\prime}}(\hat{\vec{\nu}}_{2}^{\prime}) \,
         d\hat{\vec{\nu}}_{1} d\hat{\vec{\nu}}_{2}
         d\hat{\vec{\nu}}_{1}^{\prime} d\hat{\vec{\nu}}_{2}^{\prime}
\nonumber
\end{eqnarray}
where $W(\hat{\vec{\nu}}_{1}^{\prime}, \hat{\vec{\nu}}_{2}^{\prime},t=0)$
is the probability distribution for $\hat{\vec{\nu}}_{1}^{\prime}$ and
$\hat{\vec{\nu}}_{2}^{\prime}$ at $t=0$. In the following, we will use
an isotropic distribution $W = 1/(4\pi)^{2}$ since the interaction potential
of the particles does not depend on their orientations. However, one could also
think about a situation where one first ``aligns'' the particles and then lets
them evolve with time $t$. In such a case, $W$ would be given by a product 
of delta functions. The time evolution of the correlation function is
calculated from a master equation that we derive by taking the 
time derivative of Eq.\ (\ref{11}), then using the Smoluchowski 
equation (\ref{3}) for the propagator and finally letting 
$\hat{\vec{L}}_{S} = \hat{\vec{L}}_{S}^{+}$ act on the spherical harmonics 
at time $t$:
\begin{eqnarray}
\fl
\frac{\partial}{\partial t} 
\Gamma(^{l_{1}l_{2}m_{1}m_{2}}_{l_{1}^{{\prime}}l_{2}^{\prime}
            m_{1}^{\prime}m_{2}^{\prime}} | t ) & = &
\int\int [\hat{\vec{L}}_{S}
         Y^{l_{1}*}_{m_{1}}(\hat{\vec{\nu}}_{1})
         Y^{l_{2}*}_{m_{2}}(\hat{\vec{\nu}}_{2})] \,
         P(\hat{\vec{\nu}}_{1}, \hat{\vec{\nu}}_{2},t|
   \hat{\vec{\nu}}_{1}^{\prime}, \hat{\vec{\nu}}_{2}^{\prime},0) \, 
\nonumber\\
\fl    & &  \qquad \quad\times 
         W(\hat{\vec{\nu}}_{1}^{\prime}, \hat{\vec{\nu}}_{2}^{\prime},t=0)\,
         Y^{l_{1}^{\prime}}_{m_{1}^{\prime}}(\hat{\vec{\nu}}_{1}^{\prime}) \,  
           Y^{l_{2}^{\prime}}_{m_{2}^{\prime}}(\hat{\vec{\nu}}_{2}^{\prime}) \,
         d\hat{\vec{\nu}}_{1} d\hat{\vec{\nu}}_{2}
         d\hat{\vec{\nu}}_{1}^{\prime} d\hat{\vec{\nu}}_{2}^{\prime} \enspace.
\label{12}
\end{eqnarray}
With the help of Eq. (\ref{9}) and the defintion (\ref{11}), the master
equation assumes the form
\begin{eqnarray}
\fl
-\frac{\partial}{\partial t} 
\Gamma(^{l_{1}l_{2}m_{1}m_{2}}_{l_{1}^{{\prime}}l_{2}^{\prime}
            m_{1}^{\prime}m_{2}^{\prime}} | t ) & = &
\{ D^{\perp}_{11} [l_{1}(l_{1}+1)+l_{2}(l_{2}+1)]
+ \Delta D_{11} (m_{1}^{2}+m_{2}^{2}) \nonumber \\
\fl & & \qquad \qquad + 2 D^{\|}_{12} m_{1}m_{2} \}
\Gamma(^{l_{1}l_{2}m_{1}m_{2}}_{l_{1}^{{\prime}}l_{2}^{\prime}
            m_{1}^{\prime}m_{2}^{\prime}} | t ) 
\label{13} \\
\fl & & + D^{\perp}_{12} [c^{+}_{l_{1}m_{1}} c^{-}_{l_{2}m_{2}} 
  \Gamma(^{l_{1}l_{2}m_{1}+1m_{2}-1}_{l_{1}^{{\prime}}l_{2}^{\prime}
            m_{1}^{\prime}m_{2}^{\prime}} | t )
    +                 c^{-}_{l_{1}m_{1}} c^{+}_{l_{2}m_{2}} 
  \Gamma(^{l_{1}l_{2}m_{1}-1m_{2}+1}_{l_{1}^{{\prime}}l_{2}^{\prime}
            m_{1}^{\prime}m_{2}^{\prime}} | t ) ]
\nonumber
\end{eqnarray}
The time evolution of various observables can now be calculated with
the help of this equation. We illustrate two cases which should be 
measurable in experiments.

\subsection{One-particle diffusion} \label{sec:corr.one}
\begin{figure}
\begin{indented}
\item
\includegraphics[angle=-90,width=0.6\textwidth]{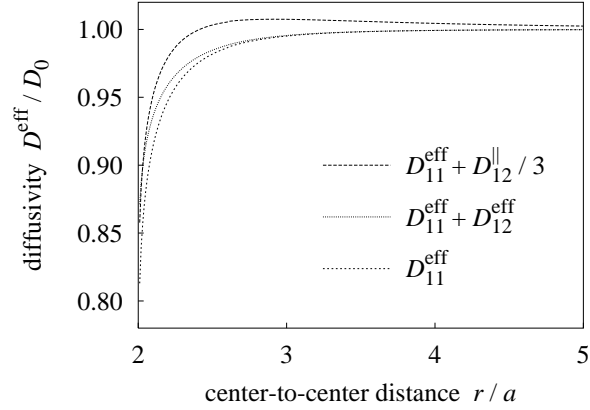}
\end{indented}
\caption{The effective rotational diffusion constants 
$D_{11}^{\mathrm{eff}}$, $D_{11}^{\mathrm{eff}}+D^{\|}_{12}/3$,
and $D_{11}^{\mathrm{eff}} + D_{12}^{\mathrm{eff}}$
in units of $D_0$ as a function of the reduced 
particle separation $r/a$.}
\label{fig:diffeff}
\end{figure}

Let us investigate the time correlation function for the orientation
of particle 1:
\begin{equation}
\langle \, \hat{\vec{\nu}}_{1}(t) \cdot \hat{\vec{\nu}}_{1}(0)\, \rangle
= \frac{4\pi}{3} \sum_{m=-1}^{1} \!
\langle \, Y^{1*}_{m}[\hat{\vec{\nu}}_{1}(t)] 
           Y^{1}_{m}[\hat{\vec{\nu}}_{1}(0)] \, \rangle
= \frac{(4\pi)^{2}}{3} \Gamma(^{10m0}_{10m0}|t)
\label{14}
\end{equation}
The first equality is just the addition theorem for spherical
harmonics\ \cite{Messiah61} and the second equality used 
$l_{2}=l_{2}^{\prime}=m_{2}=m_{2}^{\prime} = 0$, i.e., 
$Y^{0}_{0} = 1/\sqrt{4\pi}$ for the second particle in the 
definition\ (\ref{10}) of $\Gamma(\ldots|t)$. 
Since $c^{+/-}_{00} = 0$, $\Gamma(^{10m0}_{10m0}|t)$
does not couple to other $\Gamma$'s. The evolution equation is therefore
simple and gives
\begin{equation}
\Gamma(^{10m0}_{10m0}|t) = \Gamma(^{10m0}_{10m0}| 0) 
e^{-(2D_{11}^{\perp}+\Delta D_{11} m^{2})t} \enspace
\end{equation}
Compared to the single-particle result of Eq. (\ref{1a}) ($l=1$),
the decay rate of the correlation function now also depends on the
azimuthal quantum number $m$ due to presence of the second particle.
With $\Gamma(^{10m0}_{10m0}| 0) = 1/(4\pi)^2$ [isotropic distribution of
$\hat{\vec{\nu}}_{1}(0)$], the correlation function (\ref{14}) becomes
\begin{equation}
\langle \, \hat{\vec{\nu}}_{1}(t) \cdot \hat{\vec{\nu}}_{1}(0)\, \rangle
= \frac{1}{3} \, e^{-2D_{11}^{\perp}t} \Big(1+2e^{-\Delta D_{11} t}\Big)
\enspace .
\end{equation}
Finally, the mean-square angular displacement of $\hat{\vec{\nu}}_1(t)$
is calculated from
$
\langle \, |\hat{\vec{\nu}}_1(t) - \hat{\vec{\nu}}_1(0)|^{2} \, \rangle = 
2(1-\langle \, \hat{\vec{\nu}}_{1}(t) \cdot \hat{\vec{\nu}}_{1}(0)\, \rangle)
$ 
and for small times it reads
%, the mean-square angular displacement of $\hat{\vec{\nu}}_1(t)$ reads
\begin{equation}
\langle \, |\hat{\vec{\nu}}_1(t) - \hat{\vec{\nu}}_1(0)|^{2} \, \rangle
\approx 4\, D_{11}^{\mathrm{eff}} \, t \enspace \mathrm{with}
\enspace D_{11}^{\mathrm{eff}} =\frac{2D_{11}^{\perp}+D_{11}^{\|}}{3}
\enspace.
\label{14c}
\end{equation}
So $D_0$ in the analogous single-particle equation\ (\ref{1b}) is
replaced by the average of the self-diffusion constants $D_{11}^{\perp}$
and $D_{11}^{\|}$ that encode the presence of the second particle.
The effective diffusion constant $D_{11}^{\mathrm{eff}}$
in units of $D_0$ is plotted in Fig.\ \ref{fig:diffeff} as a
function of the reduced particle separation $r/a$.

\subsection{Two-particle diffusion} \label{sec:corr.two}

To access the diffusion constant $D_{12}^{\|}$, we consider the
correlation function
\begin{equation}
\langle \, \hat{\vec{\nu}}_{1}(t) \cdot \hat{\vec{\nu}}_{2}(t) \,
           \hat{\vec{\nu}}_{1}(0) \cdot \hat{\vec{\nu}}_{2}(0) \, \rangle
= \Big(\frac{4\pi}{3}\Big)^2 \sum_{m_1=-1}^{1}\sum_{m_2=-1}^{1}
   \Gamma(^{11m_1m_1}_{11m_2m_2}|t) \, .
\label{15}
\end{equation}
The correlation functions $\Gamma(^{11\pm 1\pm 1}_{11m_2m_2}|t)$
do not couple to other $\Gamma$'s and therefore relax with a rate
$2(D_{11}^{\perp}+D_{11}^{\|}+D_{12}^{\|})$ as determined from the
first two lines of Eq. (\ref{13}). The quantity 
$\Gamma(^{1100}_{11m_2m_2}|t)$ couples to 
$\Gamma(^{111-1}_{11m_2m_2}|t)$ and $\Gamma(^{11-11}_{11m_2m_2}|t)$
but can be determined straightforwardly from the three coupled evolution 
equations. So, ultimately the correlation function
(\ref{15}) appears as a sum of three exponentials. Since we are mainly 
interested in the short-time limit of Eq.\ (\ref{15}), we present
a shortcut towards the result. With the Taylor expansion
\begin{equation}
\Gamma(^{11m_1m_1}_{11m_2m_2}|t) = \frac{1}{(4\pi)^2}\delta_{m_1m_2}
-a_{m_1} t \enspace,
\end{equation}
where the coefficients $a_{m_1}$ are directly determined from the master
equation\ (\ref{13}) using 
$\Gamma(^{111-1}_{11m_2m_2}|0)=\Gamma(^{11-11}_{11m_2m_2}|0) = 0$, we
immediately arrrive at
\begin{equation}
\langle \, \hat{\vec{\nu}}_{1}(t) \cdot \hat{\vec{\nu}}_{2}(t) \,
           \hat{\vec{\nu}}_{1}(0) \cdot \hat{\vec{\nu}}_{2}(0) \, \rangle
\approx \frac{1}{3}\,[1-4(D_{11}^{\mathrm{eff}} + D_{12}^{\|}/3) t]
\enspace.
\end{equation}
Note that $D_{11}^{\mathrm{eff}}$, familiar from the one-particle diffusion
[see Eq.\ (\ref{14c})], is modified here by $D_{12}^{\|}/3$. 
Figure\ \ref{fig:diffeff} illustrates $D_{11}^{\mathrm{eff}} + D_{12}^{\|}/3$
as a function of particle separation.

In the same manner, we can also determine the short-time limit of
\begin{equation}
\langle \, \hat{\vec{\nu}}_{1}(t) \cdot \hat{\vec{\nu}}_{2}(0) \,
           \hat{\vec{\nu}}_{2}(t) \cdot \hat{\vec{\nu}}_{1}(0) \, \rangle
= \Big(\frac{4\pi}{3}\Big)^2 \sum_{m_1=-1}^{1}\sum_{m_2=-1}^{1}
   \Gamma(^{11m_1m_2}_{11m_2m_1}|t)
\end{equation}
and obtain
\begin{equation}
\langle \, \hat{\vec{\nu}}_{1}(t) \cdot \hat{\vec{\nu}}_{2}(0) \,
           \hat{\vec{\nu}}_{2}(t) \cdot \hat{\vec{\nu}}_{1}(0) \, \rangle
\approx \frac{1}{3}\,[1-4(D_{11}^{\mathrm{eff}} + D_{12}^{\mathrm{eff}}) t]
\end{equation}
with
\begin{equation}
D_{12}^{\mathrm{eff}} =\frac{2D_{12}^{\perp}+D_{12}^{\|}}{3} \enspace.
\end{equation}
We again plot $D_{11}^{\mathrm{eff}} + D_{12}^{\mathrm{eff}}$ in Fig.\ 
\ref{fig:diffeff}. A comparison of all three effective diffusion
constants in Fig.\ \ref{fig:diffeff} only reveals small differences
that pose a challenge for measurements in an experiment. Nevertheless, 
for small particle distances, the effect of hydrodynamic interactions 
calculated on the basis of lubrication theory should be measurable.

A natural extension of the theory presented here is a chain of more 
than two particles. However, the mobilites or diffusion tensors 
in such a system cannot any longer be described by the two-particle system; 
when a particle is situated between two other colloids, three-body effects 
become important, certainly for small separations.
So experiments could be used to measure deviations from the effective 
diffusion constants presented in this article. Furthermore, the coupling
to positional fluctuations of the particles has to be incorporated.

%\begin{equation}
%\vec{F}_{i} = F^{\phi}\vec{e}_{i}^{\phi}
%- K^{r}(r_{i}-R)\vec{e}_{i}^{r} \, . 
%\label{eq:vortex-force}
%\end{equation}

%\begin{figure}
%\begin{indented}
%\item
%\includegraphics[width=0.7\textwidth]{ring-trap}
%\end{indented}
%\caption{}
%\label{fig:}
%\end{figure}

%\section{Conclusion} \label{sec:conclusion}

%%% ACKNOWLEDGMENTS %%%%%%%%%%%%%%%%%%%%%%%%%%%%%%%%%%%%%%%%%%%%%%%%%%%%%%%%%%%

%\ack

%We are grateful to Erwin Frey who initiated this work by pointing out
%the experiments of Jennifer Curtis and David Grier. Furthermore, we
%would like to thank Jennifer Curtis for stimulating discussions.
%This work was supported by the Deutsche For\-schungs\-ge\-mein\-schaft
%through the Sonderforschungsbereich Transregio 6 ``Physics of colloidal
%dispersions in external fields''. H. S. acknowledges financial support
%from the Deutsche Forschungsgemeinschaft by Grant No. Sta 352/5-1.

%%% REFERENCES %%%%%%%%%%%%%%%%%%%%%%%%%%%%%%%%%%%%%%%%%%%%%%%%%%%%%%%%%%%%%%%%

\end{document}